\documentclass[11pt,twoside]{article}

\usepackage{graphicx}
\usepackage{asp2006}
\usepackage{epsf}
\usepackage{psfig}
\usepackage{lscape}
\usepackage[usenames]{}

\begin{document}
\title{The VIRMOS-VLT Deep Survey: the last 10 billion years of evolution of galaxy clustering }  
\author{A.Pollo$^{1,17}$, L. Guzzo$^{9}$, O. Le F\`evre$^{1}$, B. Meneux$^{2,9}$, A. Cappi$^{3}$, H.J. McCracken$^{10,11}$, A. Iovino$^{9}$, C. Marinoni$^{18}$, D. Bottini$^{2}$, B. Garilli$^{2}$, V. Le Brun$^{1}$, D. Maccagni$^{2}$, J.P. Picat$^{7}$, R. Scaramella$^{4,13}$, M. Scodeggio$^{2}$, L. Tresse$^{1}$, G. Vettolani$^{4}$, A. Zanichelli$^{4}$, C. Adami$^{1}$, S. Arnouts$^{1}$, S. Bardelli$^{3}$, M. Bolzonella$^{3}$, S. Charlot$^{8,10}$, P. Ciliegi$^{3}$, T. Contini$^{7}$, S. Foucaud$^{21}$, P. Franzetti$^{2}$, I. Gavignaud$^{12}$, O. Ilbert$^{20}$, B. Marano$^{6}$, A. Mazure$^{1}$, R. Merighi$^{3}$, S. Paltani$^{15,16}$, R. Pell\`o$^{7}$, L. Pozzetti$^{3}$, M. Radovich$^{5}$, G. Zamorani$^{3}$, E. Zucca$^{3}$, M. Bondi$^{4}$, A. Bongiorno$^{6}$, J. Brinchmann$^{19}$, O. Cucciati$^{9,14}$, S. de la Torre$^{1}$, F. Lamareille$^{7}$, Y. Mellier$^{10,11}$, P. Merluzzi$^{5}$, S. Temporin$^{9}$, D. Vergani$^{2}$, C.J. Walcher$^{1}$ }

\affil{$^{1}$ Laboratoire d'Astrophysique de Marseille, UMR 6110 CNRS - Universit\'e de Provence,  BP8, 13376 Marseille Cedex 12, France,\\
$^{2}$ IASF - INAF - via Bassini 15, I-20133, Milano, Italy,
$^{3}$ INAF - Osservatorio Astronomico di Bologna - Via Ranzani, 1, I-40127, Bologna, Italy,
$^{4}$ IRA - INAF - Via Gobetti,101, I-40129, Bologna, Italy,
$^{5}$ INAF - Osservatorio Astronomico di Capodimonte - Via Moiariello 16, I-80131, Napoli, Italy,
$^{6}$ Universit\`a di Bologna, Dipartimento di Astronomia - Via Ranzani,1, I-40127, Bologna, Italy,\\
$^{7}$ Laboratoire d'Astrophysique de l'Observatoire Midi-Pyr\'en\'ees (UMR 5572) - 14, avenue E. Belin, F31400 Toulouse, France,
$^{8}$ Max Planck Institut fur Astrophysik, 85741, Garching, Germany,
$^{9}$ INAF - Osservatorio Astronomico di Brera - Via Brera 28, Milan, Italy,\\
$^{10}$ Institut d'Astrophysique de Paris, UMR 7095, 98 bis Bvd Arago, 75014 Paris, France,
$^{11}$ Observatoire de Paris, LERMA, 61 Avenue de l'Observatoire, 75014 Paris, France,
$^{12}$ Astrophysical Institute Potsdam, An der Sternwarte 16, D-14482 Potsdam, Germany,
$^{13}$ INAF - Osservatorio Astronomico di Roma - Via di Frascati 33, I-00040, Monte Porzio Catone, Italy,
$^{14}$ Universit\'a di Milano-Bicocca, Dipartimento di Fisica - Piazza delle Scienze, 3, I-20126 Milano, Italy,
$^{15}$ Integral Science Data Centre, ch. d'\'Ecogia 16, CH-1290 Versoix,
$^{16}$ Geneva Observatory, ch. des Maillettes 51, CH-1290 Sauverny, Switzerland,\\
$^{17}$ Astronomical Observatory of the Jagiellonian University, ul Orla 171, 30-244 Krak{\'o}w, Poland,
$^{18}$ Centre de Physique Th\'eorique, UMR 6207 CNRS-Universit\'e de Provence, F-13288 Marseille France,
$^{19}$ Centro de Astrofísica da Universidade do Porto, Rua das Estrelas, 4150-762 Porto, Portugal, 
$^{20}$ Institute for Astronomy, 2680 Woodlawn Dr., University of Hawaii, Honolulu, Hawaii, 96822,
$^{21}$  School of Physics \& Astronomy, University of Nottingham, University Park, Nottingham, NG72RD, UK
}
\begin{abstract} 
We discuss the evolution of clustering of galaxies in the Universe from the present epoch back to $z \sim 2$, using the first-epoch data from the VIMOS-VLT Deep Survey (VVDS). We present the evolution of the projected two-point correlation function of galaxies for the global galaxy population, as well as its dependence on galaxy intrinsic luminosities and spectral types. While we do not find strong variations of the correlation function parameters with redshift for the global galaxy population, the clustering of objects with different intrinsic luminosities evolved significantly during last 8-10 billion years. Our findings indicate that bright galaxies in the past traced higher density peaks than they do now and that the shape of the correlation function of most luminous galaxies is different from observed for their local counterparts, which is a supporting evidence of a non-trivial evolution of the galaxy vs. dark matter bias. 
\end{abstract}

According to the current paradigm, galaxies formed and evolved inside dark matter halos, which merged and grew under the effect of gravity \citep{pol1:white}. Understanding the galaxy clustering may be the key to understand the evolution of the Universe itself. It is therefore an important issue of contemporary cosmology to follow the evolution of galaxy clustering in the past and to understand the bias between the galaxy distribution and the underlying dark matter density field, and how it depends on galaxy types, luminosities and local environment. 

We explored these issuees using the unique set of data from the VIMOS-VLT Deep Survey, which is strictly selected in magnitude in the range $17.5 \leq I_{AB} \leq 24$, from a complete deep photometric survey, without any color selection \citep{pol1:LEF04}. The first-epoch VVDS data contain in total 11564 galaxies with measured spectroscopic redshifts up to $z \sim 5$, making it an ideal tool to investigate the galaxy clustering in a broad range of time. Here we describe briefly our results on the evolution of the clustering properties of the general VVDS galaxy population and the dependence of clustering on galaxy luminosities, types and colors.
We measure the galaxy 2-point spatial correlation function $\xi(r_p, \pi)$ and project it along the line of sight to obtainthe function $w_p(r_p)$. The best power-law fit to $w_p(r_p)$ gives us the correlation length $r_0$ and the slope of a correlation function, $\gamma$ \citep[details in][]{pol1:techcorr}.

\section{Clustering of the general population galaxies from $z \sim 2$}

For the general population of VVDS galaxies 
we find $r_0$ roughly constant 
with a low value of $r_0=2.4_{-0.4}^{+0.4}$ $h^{-1}$ Mpc for $z=[0.2,0.5]$,
to $r_0=3.0_{-0.6}^{+0.5}$ $h^{-1}$ Mpc for $z=[1.3,2.1]$. This - seemingly surprising - lack of evolution may be understood when we take into account that we are dealing with a mixture of galaxies of different types and luminosities at different redshifts. At low z we observe mainly a low-luminosity, weakly clustered population. At high z, where we theoretically expect to observe galaxies to be less clustered than today, we probe only the most luminous, most massive and most clustered objects. 

\section{Luminosity-dependent galaxy clustering at $z \sim 1$}

\begin{figure*}[!ht]
\begin{center}
\includegraphics[width = 185pt, height = 185pt]{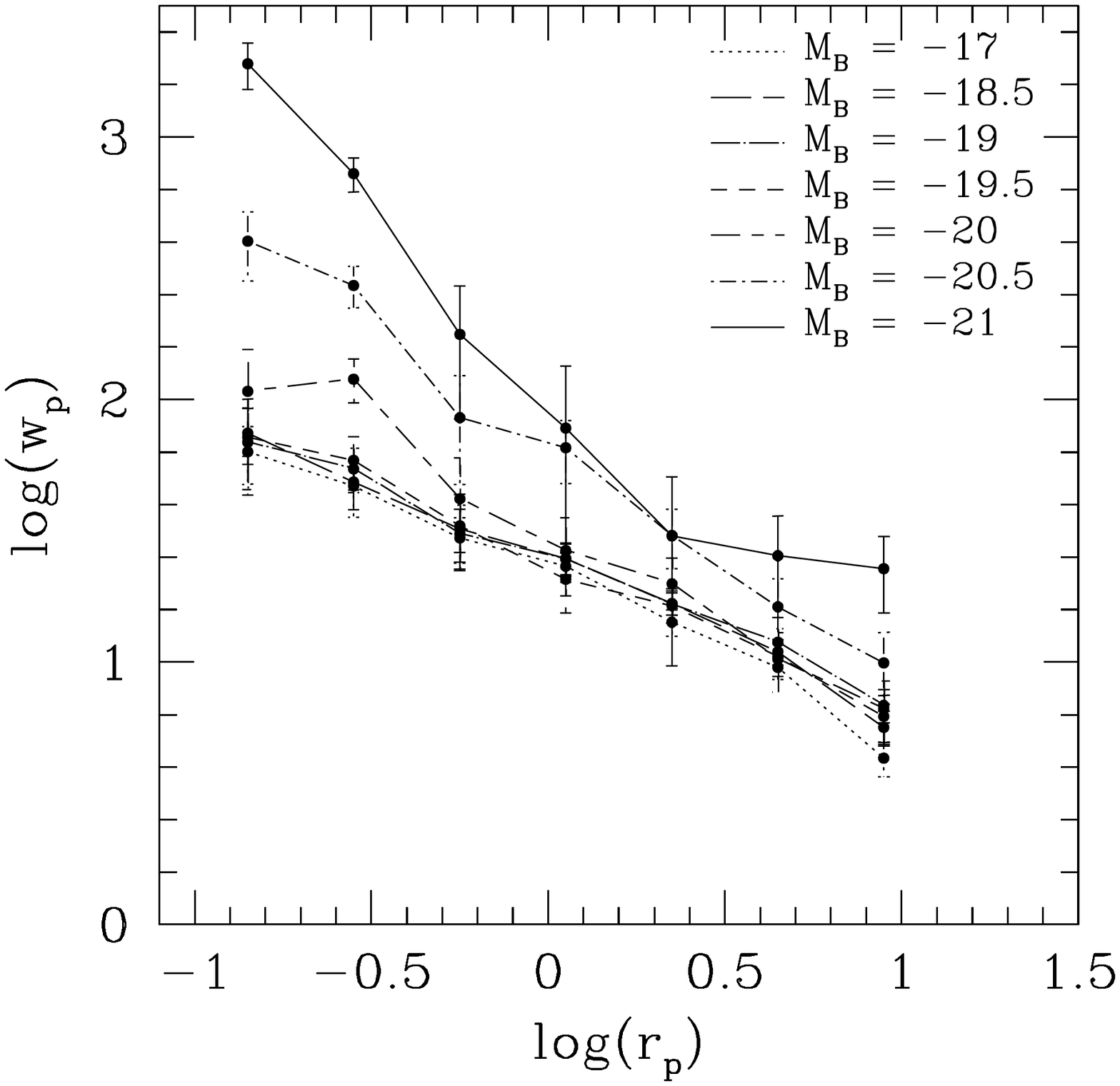}
\includegraphics[width = 185pt, height = 185pt]{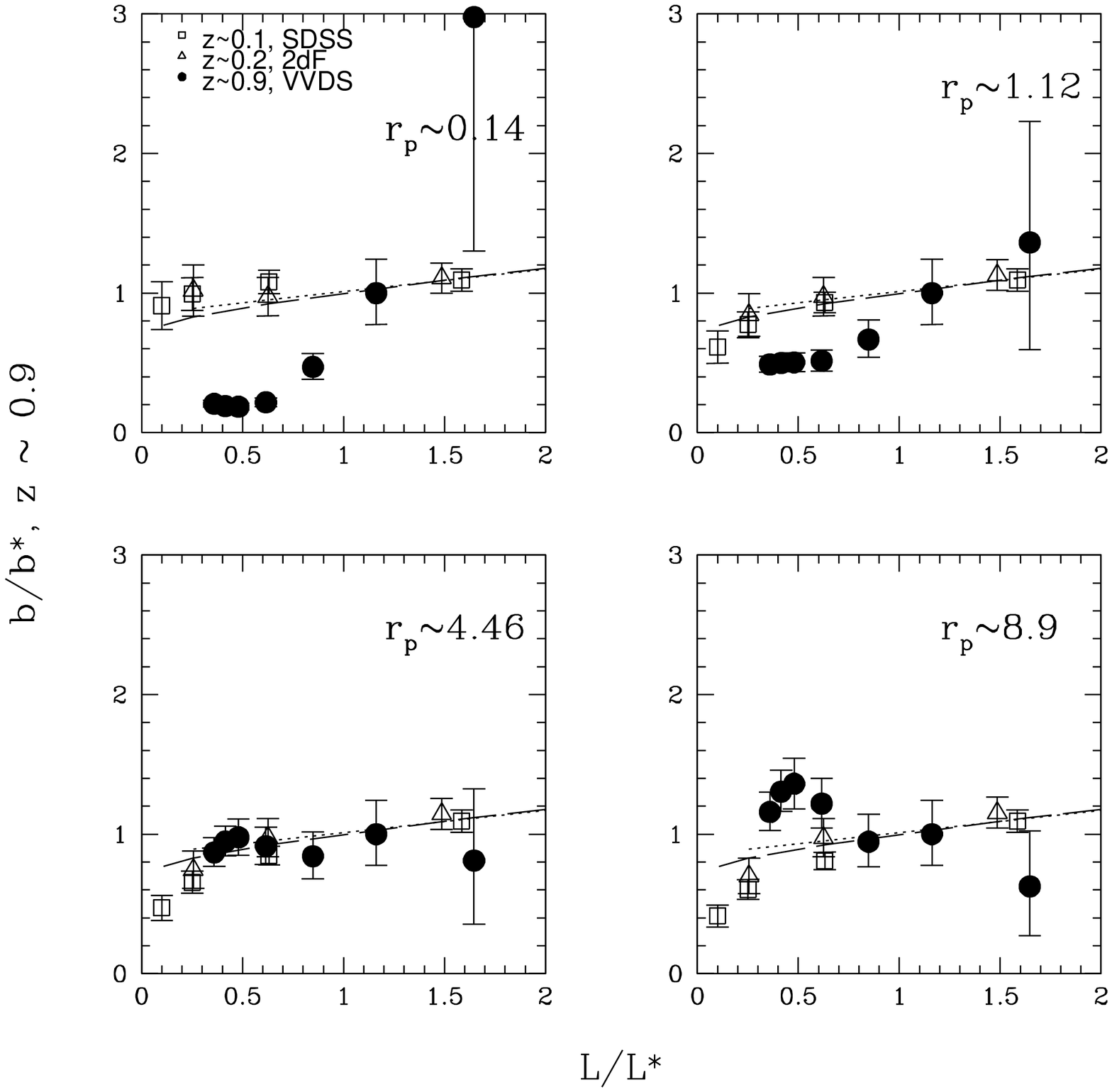}
\end{center}
\caption{Left: Projected 2-point correlation function $w_p$, measured
at $z \sim 0.9$ for volume-limited samples with limiting absolute
magnitudes up to $M_B = -21$; its shape for the brightest samples tend
to deviate from a single power-law fit, with an upturn at small scales.
Right: As a result, the relative bias strongly depends both on galaxy
intrinsic luminosities and on the scale on which it is measured,
unlike observed locally, e.g. in SDSS \citep{pol1:zehavi} and 2dF \citep{pol1:norberg02}.}\label{pol1:fig1}
\end{figure*}

Luminosity-dependence of galaxy clustering at $z \sim 1$ is clearly different with what is observed in the nearby universe. The clustering strength is rising around the characteristic Schechter luminosity $M_B^*$, with a sharper turn than observed at low redshifts. Additionally, the slope of the correlation function steepens from $\gamma=1.6^{+0.1}_{-0.1}$ to $\gamma=2.4^{+0.4}_{-0.2}$. This is caused by a significant change in the shape of $w_p(r_p)$, increasingly deviating from a power-law for the most luminous samples, with a strong upturn at small ($\le 1-2$ $h^{-1}$ Mpc) scales (see left panel of Figure \ref{pol1:fig1}). This trend, not observed locally, results also in a visible scale-dependence of the relative bias, $b/b*$ (shown in the right panel of Figure \ref{pol1:fig1}) and seems to imply a significant change in the way luminous galaxies trace dark-matter halos at $z\sim 1$ with respect to the present-day Universe.

\section{Evolution of clustering of galaxies of different types and colors}

\begin{figure*}[!ht]
\begin{center}
\includegraphics[width = 185pt, height = 185pt]{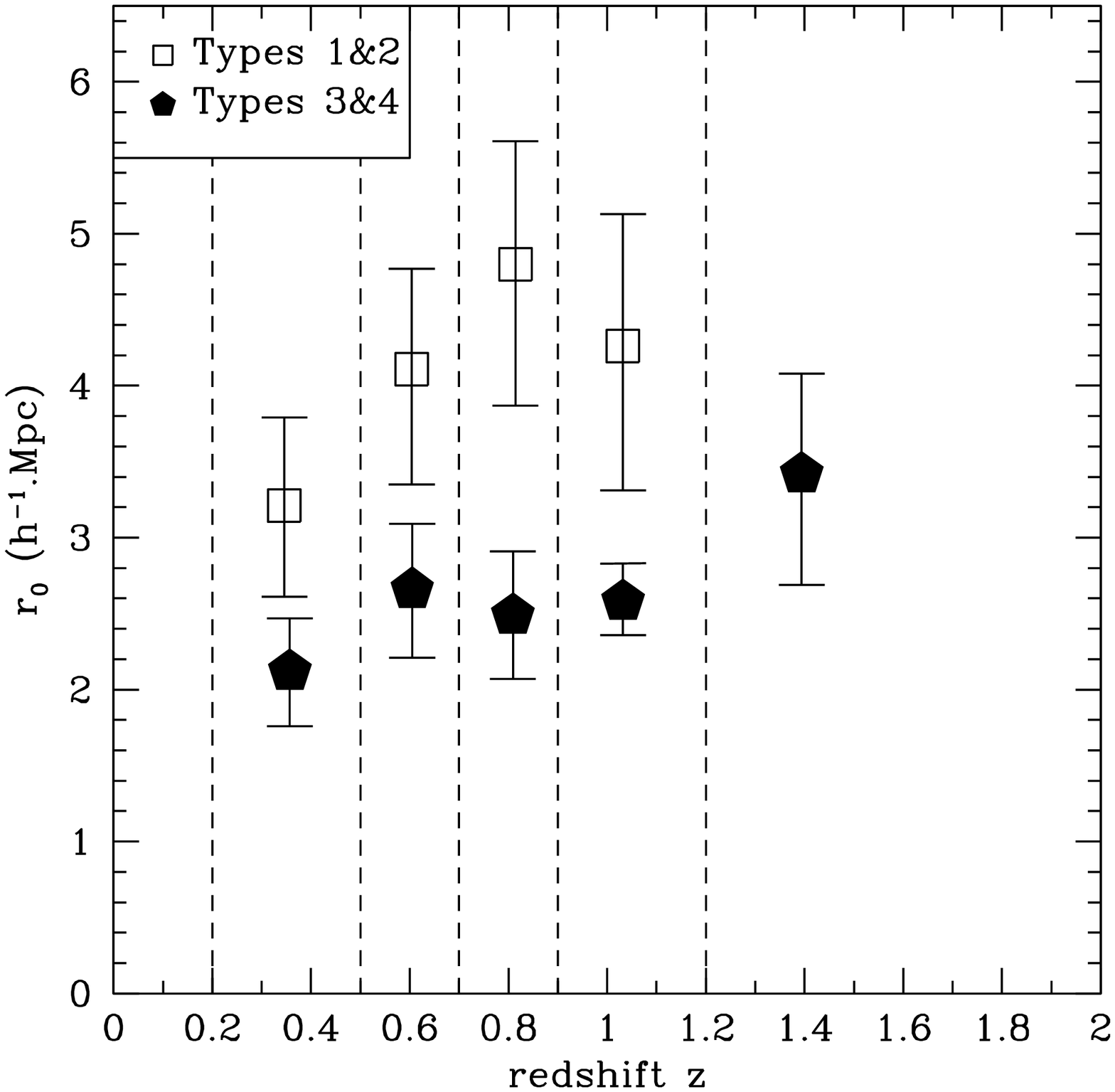}
\includegraphics[width = 185pt, height = 185pt]{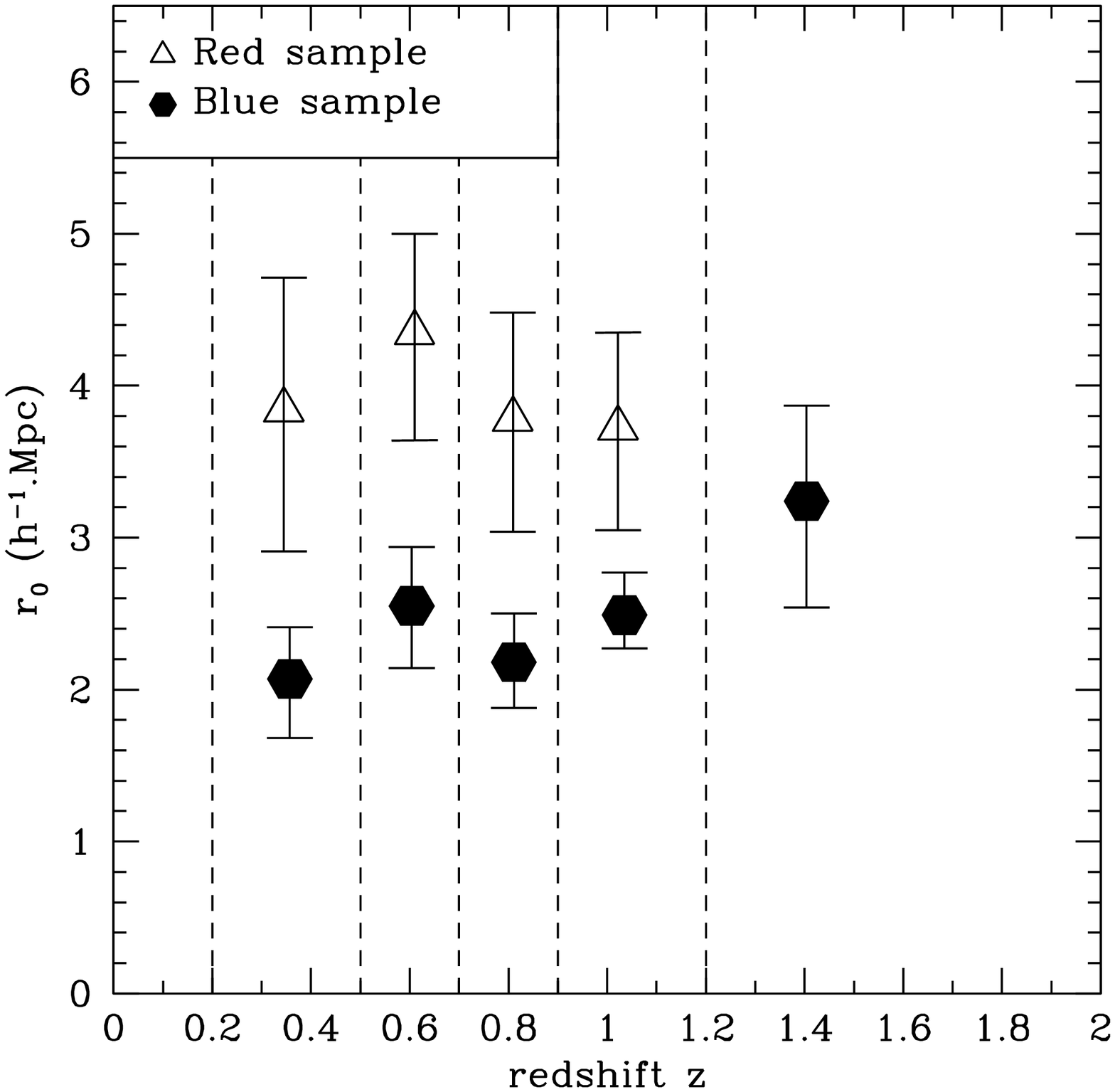}
\end{center}
\caption{Left: Correlation length $r_0$ as a function of redshift for early (T1\&2) and late (T3\&4) type galaxies. Right: Correlation length $r_0$ as a function of redshift for the red and blue (in rest-frame) galaxies.
}\label{pol1:fig2}
\end{figure*}

As shown in the left panel of Figure \ref{pol1:fig2}, the clustering of galaxies of early spectral types is stronger than that of late-type galaxies at all redshifts up to $z\simeq1.2$.  At $z\sim 0.8 $, early-type galaxies display a correlation length $r_0=4.8 \pm$0.9~h$^{-1}$Mpc, while late types have $r_0=2.5 \pm $0.4~h$^{-1}$Mpc. The relative bias between early- and late-type galaxies remains approximately constant with $b=1.6 \pm 0.3$ from $z=0$ to $z=1.2$. Similarly, as shown in the right panel of Figure \ref{pol1:fig2}, red-sequence galaxies exhibit a larger clustering length than the blue ones with little dependence on redshift. These results \citep{pol1:meneux} are consistent with a scenario where the assembly of the most luminous early-type galaxies is practically completed at $z \sim 1$, with however luminous star-forming objects being nearly as clustered as red galaxies for $z$ larger than $\sim 1.5$.  

All the above-mentioned measurements are pieces of a puzzle which still have to be fit into the theoretical scheme of gravitational instability scenario, where dark matter halos assemble under the effect of gravity and galaxies form and evolve inside them. Our observations represent an important constraint for models trying to reproduce galaxy clustering, especially at small, non-linear scales.

\begin{acknowledgements}
This research has been developed within the framework of the VVDS
consortium.
This work has been partially supported by the
CNRS-INSU and its Programme National de Cosmologie (France),
and by Italian Ministry (MIUR) grants
COFIN2000 (MM02037133) and COFIN2003 (num.2003020150).
The VLT-VIMOS observations have been carried out on guaranteed
time (GTO) allocated by the European Southern Observatory (ESO)
to the VIRMOS consortium, under a contractual agreement between the
Centre National de la Recherche Scientifique of France, heading
a consortium of French and Italian institutes, and ESO,
to design, manufacture and test the VIMOS instrument.
\end{acknowledgements}

\end{document}